# Predicting accurate absolute binding energies in aqueous solution: thermodynamic considerations for electronic structure methods

**Jan H. Jensen**
Department of Chemistry
University of Copenhagen
Universitetsparken 5
2100 Copenhagen, Denmark
Email: jhjensen@chem.ku.dk, Twitter @janhjensen

**Abstract**
Recent predictions of absolute binding free energies of host-guest complexes in aqueous solution using electronic structure theory have been encouraging for some systems, while other systems remain problematic. In this paper I summarize some of the many factors that could easily contribute 1-3 kcal/mol errors at 298 K: three-body dispersion effects, molecular symmetry, anharmonicity, spurious imaginary frequencies, insufficient conformational sampling, wrong or changing ionization states, errors in the solvation free energy of ions, and explicit solvent (and ion) effects that are not well-represented by continuum models. While I focus on binding free energies in aqueous solution the approach also applies (with minor adjustments) to any free energy difference such as conformational or reaction free energy differences or activation free energies in any solvent.

**Introduction**
The prediction of accurate absolute binding energies in aqueous solution is one of the holy grails of computational chemistry, mainly because of the potential use in rational drug design. "Accurate" is typically taken to be 1 kcal/mol, which roughly corresponds to predicting the binding constant within an order of magnitude at room temperature and it is understood that the method must be generally applicable. The recent blind prediction challenge SAMPL4 has shown that this goal has yet to be met even for host-guest complexes that are significantly smaller than proteins (Muddana *et al.* 2014). Interestingly, the entry that arguably performed best for one of the hosts (curcurbit[7]uril or CB7) was, for the first time, based on the rigid rotor-harmonic oscillator (RRHO) approximation and electronic structure theory and involved no direct parameterization against experimental binding free energies (Sure *et al*. 2014). This method reproduced 14 experimental CB7-guest binding free energies with a mean absolute deviation of 2.02 ± 0.46 kcal/mol suggesting that, perhaps, the holy grail is within reach. However, the mean absolute error was significantly larger for another host-guest system indicating that there remains some work to be done.

In this paper I summarize why electronic structure/RRHO-based approaches are starting to yield accurate binding free energies. I also discuss many of the possible sources of error when computing aqueous binding free energies with electronic structure theory and how to correct for them.

**General approach**
The general approach for predicting the standard free energy of binding ($\Delta G^o_{b,aq}$) of a receptor (R or host) and ligand (L or guest) molecule in aqueous (aq) solution

$$R(aq) + L(aq) \rightleftharpoons RL(aq) \qquad (Rx\ 1)$$

using electronic structure theories is through a thermodynamic cycle (Figure 1)

$$\Delta G^o_{b,aq} = G^o_{aq}(RL) - G^o_{aq}(R) - G^o_{aq}(L) \qquad (1)$$

where

$$G^o_{aq}(X) = E_{gas}(X) + G^o_{gas,RRHO}(X) + \Delta G^o_{solv}(X)$$
$$= G^o_{gas}(X) + \Delta G^o_{solv}(X) \qquad (2)$$

$E_{gas}(X)$, $G^o_{gas,RRHO}(X)$, and $\Delta G^o_{solv}(X)$ is the electronic energy, rigid rotor-harmonic oscillator (RRHO), and solvation free energy, respectively, of molecule X. Note that $G^o_{gas,RRHO}(X)$ contains the zero point energy. The standard state (denoted by "o") throughout this paper is 1 M, unless otherwise noted. The solvation free energy is typically computed using a continuum solvation model as described in detail below.

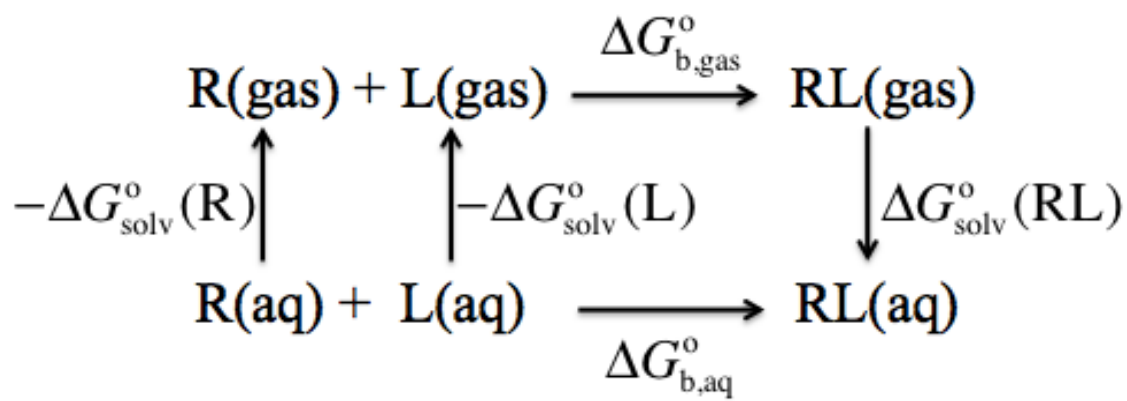

**Figure 1.** Thermodynamic cycle for computing the binding free energy in aqueous solution for a ligand (L) binding to a receptor (R) to form a complex (RL).

**The Electronic Energy**
One of the reasons electronic structure-based approaches are starting to yield accurate binding free energies is the use of dispersion corrections (Grimme *et al*. 2010) in the evaluation of the electronic energy and the structure (as well as the vibrational frequencies as discussed below). Grimme (2012) has shown that dispersion typically makes a very big (>10 kcal/mol) contribution to binding free energies of host-guest complexes. Dispersion corrections are therefore a must if DFT is used to compute the electronic binding energy. Furthermore, Grimme has shown that three-body dispersion makes a non-negligible (2-3 kcal/mol) contribution to the electronic binding energy. For convergent methods this effect is only included in rather expensive methods that involve triple-excitations such as MP4 and CCSD(T).

Interestingly, it has been found that dispersion corrected, and short-range corrected, semiempirical methods such as DFTB or PM6, yield binding energies with accuracies similar to conventional DFT calculations with large basis sets. For example, Muddana and Gilson (2012) used PM6-DH+ to compute reasonably accurate relative binding energies for CB7-ligand complexes. On the other hand, Yilmazer and Korth (2013) found significant deviations for PM6-DH+ and similar methods when applied to larger protein-ligand models. Whether these minimal basis set-based methods are sufficiently flexible to handle large many-body polarization effects involving many charged groups remains to be determined. In any case, Grimme and co-worker have computed $E_{gas}(X)$ at the PW6B95-D3(BJ)/def2- QZVP//TPSS27-D3(BJ)/def2-TZVP level of theory with good results (Sure *et al*. 2014).

**Molecular Thermodynamics**
The translational, rotational and vibrational thermodynamic contribution to the binding free energy is very large (>10 kcal/mol) and must be included for accurate results. Some years ago there was a bit of confusion in the literature about whether the RRHO approach was appropriate for condensed phase systems, but Zhou and Gilson (2009) have clarified this beautifully. The accuracy of the dispersion and hydrogen bond-corrected semi-empirical methods mentioned above has now made it feasible to compute the vibrational frequencies for typical host-guest complexes and this is another reason why electronic structure-based approaches are starting to yield accurate binding free energies. (They appear to be a qualitative step forward in accuracy compared to standard force fields in this regard.) For example, Grimme has computed $G^\text{o}_\text{gas,RRHO}(X)$ with PM6-D3H (Grimme 2013) and HF-3c (Sure *et al.* 2014) with good results.

**The standard state**. Most electronic structure codes compute the RRHO energy corrections for an ideal gas, where the standard state is a pressure of 1 bar. As I'll discuss further below the solvation free energies are computed for a 1 M standard state so the gas phase free energy must be corrected accordingly

$$G^\text{o}_\text{gas,RRHO}(X) = G^\text{o (1 bar)}_\text{gas,RRHO}(X) - RT\ln(V^{-1}) \quad (3)$$

where $V$ is the volume of an ideal gas a temperature $T$ and $R$ is the ideal gas constant. At 298K this correction increases the free energy by 1.90 kcal/mol.

It is tempting to argue that since the volume change in solution is negligible one should use the Helmholtz free energy $A^\text{o}_\text{gas,RRHO}(X)$ instead of the Gibbs free energy. However, as I discuss below, the solvation free energy corrects for the change in volume on going from the gas phase to solution, so the Gibbs free energy change should be used throughout.

**The vibrational enthalpy for NDDO based semiempirical methods**. NDDO based semiempirical methods such as PM6 are parameterized against experimental standard enthalpies of formation ($\Delta H^\text{o}_\text{f,gas}$). However, in the case of intermolecular interactions such as hydrogen binding the parameterization was done by fitting $\Delta\Delta H^\text{o}_\text{f,gas}$ to $\Delta E_\text{gas}$ values computed using electronic structure theory (Stewart 2007). The same is true for dispersion and hydrogen bond corrected PM6 methods. Thus, if a PM6 based method is used to compute the interaction energy the RRHO enthalpy corrections should still be included, i.e.

$$G^\text{o}_\text{aq}(X) = \Delta H^\text{o}_\text{f,gas}(X) + G^\text{o}_\text{gas,RRHO}(X) + \Delta G^\text{o}_\text{solv}(X) \quad (4)$$

**Molecular symmetry**. Many host molecules and some guest molecules are symmetric and this affects the rigid-rotor rotational entropy ($S_{RR}$) through the symmetry number (σ), which is a function of the molecular point group.

$$S_{RR} = R\ln\left(\frac{8\pi^2}{\sigma}\left(\frac{2\pi ekT}{h^2}\right)^{3/2}\sqrt{I_1 I_2 I_3}\right) \quad (5)$$

Here $h$ and $k$ are Planck's and Boltzmann's constant, respective and $I_x$ is the moment of inertia for principal axis $x$. In practice it can be very difficult to build large molecules with the correct point group and most studies use $C_1$ symmetry. In this case the effect of symmetry must be added manually to the free energy

$$G^{\circ}_{\text{gas,RRHO}}(X) = G^{\circ (C_1)}_{\text{gas,RRHO}}(X) + RT \ln(\sigma_X) \quad (6)$$

As an example, CB7 has $D_{7h}$ symmetry and a corresponding σ value of 14, in which case the correction contributes 1.56 kcal/mol to the free energy at 298K.

**Anharmonicity and low frequency modes**. Host-guest complexes can exhibit very low frequency vibrations on the order of 50 cm$^{-1}$ or less, which tend to dominate the vibrational entropy contribution (Grimme 2013). Many researchers have questioned whether the harmonic approximation is valid for such low frequency modes and this is an open research question. The main problem is that it is very difficult to compute the vibrational entropy exactly. Most methods for computing anharmonic effects are developed to obtain the 1 or 2 lowest energy states, but for very low frequency modes 10-20 states are likely significantly populated at room temperature and therefore contribute to the entropy.

In the absence of theoretical benchmarks, comparison to experiment can prove constructive. Kjærgaard and co-workers (Bork *et al*. 2014a and 2014b) have recently measured standard binding free energies for small gas phase compounds and compared them to CCSD(T)/aug-cc-pV(T+d) calculations. For example, in the case of acetonitrile-HCl the measured binding free energy at 295K is between 1.2 and 1.9 kcal/mol, while the predicted value is 1.9 kcal/mol using the harmonic approximation (Bork *et al*. 2014b). Since the errors in ΔE and the rigid-rotor approximation presumably are quite low, this suggest and error in the vibrational free energy of at most 0.7 kcal/mol, despite the fact that the lowest vibrational frequency is only about 30 cm$^{-1}$. Furthermore, the error can be reduced by 0.4 kcal/mol by scaling the harmonic frequencies by anharmonic scaling factors suggested by Shields and co-workers (Temelso et al. 2011; Temelso & Shields 2011). Similar results were found for dimethylsulfide-HCl (Bork *et al*. 2014a). So there are some indications that the harmonic approximation yields free energy corrections that are reasonable and possibly can be improved upon by relatively minor corrections.

On the other hand in a recent study Piccini and Sauer (2014) show that anharmonic effects need to be included to obtain agreement with the experimental binding free energy of methane to H-CHA zeolite. Specifically, they compute the vibrational binding free energy by computing the 1-dimensional potential energy surface for each low frequency mode and compute the vibrational energy levels and corresponding partition function numerically (as opposed to using the anharmonic fundamental frequency together with the harmonic oscillator partition function). This decreases the binding free energy by 2.5 kcal/mol compared to the standard harmonic oscillator treatment.

Grimme (2012) has taken a different approach by arguing that low-frequency modes resemble free rotations and using the corresponding entropy term for low frequency modes. This changes the RRHO free energy correction by 0.5 - 4 kcal/mol, depending on the system.

Low frequencies are especially susceptible to numerical error and it is not unusual to see 1 or 2 imaginary frequencies of low magnitude in a vibrational analysis of a host-guest complex. Since imaginary frequencies are excluded from the vibrational free energy this effectively removes 1 or 2 low frequency contributions to the vibrational free energy. For example, a 30 cm$^{-1}$ frequency contributes about 1.7 kcal/mol to the free energy at 298K.

Imaginary frequencies resulting from a flat PES and numerical errors can often be removed by making the convergence criteria for the geometry optimization and electronic energy minimization more stringent and making the grid size finer in the case of DFT calculations. If the Hessian is computed using finite difference it is important to use central-differencing. If all else fails, it is probably better to pretend that the imaginary frequency is real and add the corresponding vibrational free energy contribution. However, this needs to be systematically tested.

**Conformations**. One of the main problems in computing accurate binding free energies is to identify the structures of the host, guest and (especially) the host-guest complex with the lowest free energy. Because both the RRHO and solvation energy contributions contribute greatly to the binding free energy change, simply finding the structure with the lowest electronic energy and computing the free energy only for that conformation is unlike to result in the global free energy minimum.

For a molecule (X) with $N_{\text{conf}}$ conformations the standard free energy is

$$G^{\text{o}}_{\text{aq}}(X) = G^{\text{o}}_{\text{aq}}(X_{\text{ref}}) - RT \ln\left(1 + \sum_{\substack{i=1 \\ i \neq \text{ref}}}^{N_{\text{conf}}-1} e^{-\Delta G^{\text{o}}_{\text{aq}}(X_i)/RT}\right) \qquad (7)$$

where

$$\Delta G^{\text{o}}_{\text{aq}}(X_i) = G^{\text{o}}_{\text{aq}}(X_i) - G^{\text{o}}_{\text{aq}}(X_{\text{ref}}) \qquad (8)$$

and where $X_{\text{ref}}$ is some arbitrarily chosen reference geometry - for example the global minimum. With that choice for $X_{\text{ref}}$, conformations with free energies higher than 1.36 kcal/mol contribute less than 0.1 to the sum at 298K. So a significant number of very low free energy structures are needed to make even a 0.5 kcal/mol contribution to the free energy. Conformations related by symmetry should not be included here as their effects are accounted for in the rotational entropy (see above). Note that if the binding measurements are done for racemic mixtures then all stereoisomers must be included in the sum.

**Molecular charge and pH**
Virtually all binding measurements in aqueous solution are performed in a buffer with a constant pH and many ligands and or receptors contain one or more ionizable groups. The charge ($q$) of an ionizable (acid/base) group in aqueous solution depends on its p$K_a$ and the pH:

$$q = \frac{1}{1+10^{\text{pH}-\text{p}K_a}} - \delta \quad (9)$$

where $\delta$ is 1 for an acid and 0 for a base. This is an average charge for all the molecules in solution and will not be an integer. This section describes how to handle charges that different significantly from an integer value and/or change as a result of binding. The p$K_a$ can be computed using electronic structure theory or empirically using software such as Marvin (2014). However, if the p$K_a$ value is perturbed by the binding the situation may be complicated further. Here I illustrate this point for a simple example where the ligand has a basic group that is neutral when deprotonated and the receptor is non-ionizable.

$$R(aq) + L(H^+)(aq) \rightleftharpoons RL(H^+)(aq) \quad \text{Rx 2}$$

The apparent equilibrium constant is then (throughout this paper I assume ideal solutions where the activity is equal to the concentration)

$$K' = \frac{[RL]+[RLH^+]}{[R]([L]+[LH^+])} \quad (10)$$

and the corresponding binding free energy is

$$\Delta G'^o_{aq} = \Delta G^o_{aq}(+) - RT \ln\left(\frac{1+10^{\text{pH}-\text{p}K_a^c}}{1+10^{\text{pH}-\text{p}K_a^f}}\right) = \Delta G^o_{aq}(0) - RT \ln\left(\frac{1+10^{\text{p}K_a^c-\text{pH}}}{1+10^{\text{p}K_a^f-\text{pH}}}\right) \quad (11)$$

where $\Delta G^o_{aq}(+)$ and $\Delta G^o_{aq}(0)$ is the binding free energy computed using the charged (protonated) and neutral form of the ligand and p$K_a^c$ and p$K_a^f$ are the p$K_a$ values the ligand bound to the receptor and the free ligand, respectively.

For example, Koner et al. (2011; Kim *et al.* 2014) have shown that binding of benzimidazole and derivatives to CB7 can increase the p$K_a$ of the ligand by as much as 4 pH units (from p$K_a^f$ = 4.6 and p$K_a^c$ = 8.6) which results in a 3.3 kcal/mol pH-dependent correction to the binding free energy at pH 7. Put another way, using p$K_a^f$ to determine the protonation state of the bound ligand would result in an 3.3 kcal/mol error in the binding free energy.

For many ligands of interest the p$K_a^f$ can be estimated fairly accurately in a matter of second using programs such as Marvin. The effect of binding on p$K_a^f$ can often be estimated by chemical intuition since hydrogen bonds to charged acid and basic groups tend to, respectively, lower or raise the p$K_a$ even further. For example, if an amine with p$K_a^f$ = 9 binds to the receptor via hydrogen bonding, then p$K_a^c$ is likely higher than 9 and $\Delta G'^o_{aq} \approx \Delta G^o_{aq}(+)$ is a good approximation. However, if p$K_a^f$ is close to 7 then p$K_a^c$ should be computed. Also, it is possible for charged ligands to change to their neutral state if they bind to hydrophobic or similarly charged receptors.

If p$K_a^f$ is known with some degree of confidence (e.g. from experiment or Marvin) then p$K_a^c$ can be estimated by

$$\mathrm{p}K_a^c = \mathrm{p}K_a^f - \frac{\Delta G^o_{\Delta \mathrm{p}K,\mathrm{aq}}}{RT \ln(10)} \quad (12)$$

where $\Delta G^o_{\Delta \mathrm{p}K,\mathrm{aq}}$ is the free energy change for this reaction (Li *et al.* 2014)

$$\mathrm{RLH}^+(\mathrm{aq}) + \mathrm{L}(\mathrm{aq}) \rightleftharpoons \mathrm{RL}(\mathrm{aq}) + \mathrm{LH}^+(\mathrm{aq}) \quad \mathrm{Rx\ 3}$$

However, if one suspects that empirical p$K_a$ predictors such as Marvin give inaccurate results for p$K_a^f$ then this value can be computed using quantum chemistry. Ho and Coote (2010) have written a very useful summary of different approaches to such predictions. The accuracy for phenol and carboxyl p$K_a$ values are as low at 1 pH units (unfortunately they did not give a value for amines). However, if the pKa value is close to the pH of interest a 1 pH unit-error can lead to prediction of the wrong protonation and result in errors in the binding free energy on the order of 1-3 kcal/mol.

If there are several ($N_{\mathrm{ionz}}$) ionizable groups then Eq (11) generalizes to

$$\Delta G'^o_{\mathrm{aq}} = \Delta G^o_{\mathrm{aq}}(-/+) - RT \ln \left( \sum_{i=1}^{N_{\mathrm{ionz}}} \frac{1 + 10^{s_i(\mathrm{pH} - \mathrm{p}K_{a,i}^c)}}{1 + 10^{s_i(\mathrm{pH} - \mathrm{p}K_{a,i}^f)}} \right) \quad (13)$$

where $\Delta G^o_{\mathrm{aq}}(-/+)$ is the binding free energy when all acids and bases are deprotonated and protonated, respectively, the sum runs over all ionizable groups and $s_i$ is 1 and -1 if $i$ is a base or acid, respectively.

However, this assumes that the ionizable groups titrate independently of one another, i.e. that the p$K_a$ value of one group is independent of the protonation states of all other ionizable groups. If that is not the case then it is difficult to give a general expression for the pH-dependent free energy correction in terms of p$K_a$ values (though it can be derived for a specific case). Next I present an alternative approach, but note that in practice because one can obtain more accurate *relative* p$K_a$ values (using Eq (12) or similar (Ho and Coote 2010)) than absolute p$K_a$ values it may be worth the extra effort to derive the pH-dependent free energy correction in terms of p$K_a$ values.

**Legendre transformed free energies**. Instead a general expression can be written in terms of Legendre transformed free energies as suggested by Alberty (Alberty 2005, Alberty *et al.* 2011) and modified here to electronic structure calculations (Jinich *et al.* 2014):

$$G''^o_{aq}(\bar{X}) = -RT \ln\left(\sum_{i=1}^{2^{N_{ionz}}} e^{-G''^o_{aq}(X_i)/RT}\right) \quad (14)$$

where $\bar{X}$ denotes an average over several protonation states of X, $2^{N_{ionz}}$ is the number of possible protonation states given $N_{ionz}$ sites and

$$G''^o_{aq}(X_i) = G^o_{aq}(X_i) - n_i(H^+)\left(\Delta G^o_{solv}(H^+) - RT\ln(10)pH\right) \quad (15)$$

where $n_i(H^+)$ is the number of ionizable protons in protonation state $i$, and $\Delta G^o_{solv}(H^+)$ is the solvation free energy of the proton. So in the case of ligand L considered above, $n_i(H^+)$ is 0 and 1 for L and LH$^+$, respectively.

$\Delta G^o_{solv}(H^+)$ is usually taken from the literature where estimates vary between -264 and -266 kcal/mol (Kelly *et al.* 2006), which can add to the uncertainty in the predicted binding free energy change. There are at least two ways of reducing the error. One way is to maximize error cancelation by computing $\Delta G^o_{solv}(H^+)$ (using explicit solvent molecules as discussed below) using the same level of theory method use to compute $\Delta G^o_{b,aq}$. The other way is to choose the value of $\Delta G^o_{solv}(H^+)$ used as reference for the experimental solvation free energies of ions that are used to parameterize the continuum solvation model you use (Table 1). The first way is best if explicit solvent molecules are used to compute the solvation free energies of ions in the binding study and otherwise the second method is best.

Using Legendre transformed free energies, Eq (1) can be rewritten as

$$\Delta G''^o_{b,aq} = G''^o_{aq}(R\bar{L}) - G^o_{aq}(R) - G''^o_{aq}(\bar{L}) \quad (16)$$

Since the electronic energy contribution to the standard free energy can be very large in magnitude this form is more easily evaluated

$$G''^o_{aq}(\bar{X}) = G''^o_{aq}(X_{ref}) - RT \ln\left(1 + \sum_{\substack{i=1 \\ i \neq ref}}^{2^{N_{ionz}}-1} e^{-\Delta G''^o_{aq}(X_i)/RT}\right) \quad (17)$$

where

$$\Delta G''^o_{aq}(X_i) = G''^o_{aq}(X_i) - G''^o_{aq}(X_{ref}) \quad (18)$$

and where $X_{ref}$ is some arbitrarily chosen reference protonation state, for example that for which $n_i(H^+) = 0$. The sum can be combined with that over different conformations [Eq (7)] as discussed below.

**Other ions and Ionic Strength**. If the ligand and/or hosts contain ionizable groups then the binding measurements were likely performed in a buffer, with a certain ionic strength, to regulate pH. It is possible to include this effect in continuum solvation models such as the PCM method (Cossi et al 1998). However, given the relatively low (10-100 mM) concentrations usually used in the experiments this will only have a noticeable (> 0.5 kcal/mol) effect on the energetics involving multiply charged ions. As discussed below, the error in the computed solvation energy for such ions are already large and it is not clear whether it is worth including non-specific ionic strength effects in the computations. At high ion concentrations, it is possible that these ions bind at certain sites in the ligand, receptor, or ligand-receptor complex with sufficient probability that they must be included in the thermodynamics. If so the exact same equations and considerations outlined above for $H^+$ also apply to, e.g. $Cl^-$ and $pCl^-$ (computed from the specified buffer concentration) is used instead of pH.

**Solvation thermodynamics**
**Background**. Most continuum models (CMs) of solvation compute the solvation free energy as the difference between the free energy in solution ($G_{soln,E}^{o,CM}(X)$) and the gas phase electronic energy ($E_{gas}(X)$)

$$\Delta G_{solv}^o(X) = G_{soln,E}^{o,CM}(X) - E_{gas}(X) \quad (19)$$

$G_{soln,E}^{o,CM}(X)$ typically contains energy terms describing the electrostatic interaction of the molecule and the continuum as well as the van der Waals interactions with the solvent and free energy required to create the molecular cavity in the solvent (cavitation). The electrostatic interaction with the solvent alters the molecular wavefunction and is computed self-consistently. Usually the gas phase structure of X is used for the computation of $G_{soln,E}^{o,CM}(X)$, though for COSMO-RS the structure is optimized in solution. There is typically no explicit RRHO contribution for $G_{soln,E}^{o,CM}(X)$ so the computational cost is comparable to that for $E_{gas}(X)$.

Some software packages automatically compute $\Delta G_{soln,E}^{o,CM}(X)$ and $E_{gas}(X)$ in one run, while other packages only compute $G_{soln,E}^{o,CM}(X)$. Also, some programs just compute the electrostatic component of $G_{soln,E}^{o,CM}(X)$ by default. However, the van der Waals and, especially, the cavitation component can make sizable contributions to the binding free energy and must be included for accurate results. It is worth noting that any hydrophobic contribution to binding will derive primarily from the change in cavitation energy (Cramer & Truhlar 1991 and references therein).

$G_{soln,E}^{o,CM}(X)$ contains parameters (e.g. atomic radii) that are adjusted to reproduce experimentally measured solvation free energies

$$\Delta G_{solv}^{o,exp}(X) = G_{soln}^{o,exp}(X) - G_{gas}^{o,exp}(X) \quad (20)$$

The standard state for both $G_{soln}^{o,exp}(X)$ and $G_{gas}^{o,exp}(X)$ is generally chosen to 1 M (Ben-Naim 1978; Ben-Naim & Marcus 1984). The latter is the reason a 1 M reference state also must be used when computing $G_{gas,RRHO}^{o}(X)$.

Notice that the volume on going from the gas phase to solution is included in the solvation free energy

$$\Delta G_{solv}^{o,exp}(X) = \Delta A_{solv}^{o,exp}(X) + p^{o}(\Delta V_{solv} - V_{gas}) \quad (21)$$

where $\Delta V_{solv}$ is the volume change in solution due to addition of the solute X to the neat solvent. For an ideal gas ($p^{o}V_{gas} = RT$) it follows that

$$\Delta\Delta G_{solv}^{o} = \Delta\Delta A_{solv}^{o} + p^{o}\Delta\Delta V_{solv} - RT \quad (22)$$

and

$$\Delta G_{b,aq}^{o} = \Delta A_{b,aq}^{o} + p^{o}\Delta V_{soln} \quad (23)$$

because the $-RT$ term is cancelled by a corresponding term in the translational enthalpy contribution to $\Delta G_{gas,RRHO}^{o}$. $\Delta V_{soln} = \Delta\Delta V_{solv}$ is the change in the volume of the solution on upon binding.

**Atomic radii**. The solvation energy is computed using a set of atomic radii that define the solute-solvent boundary surface. These radii are usually obtained by fitting to experimentally measured solvation energies. Accurate solvation energies should not be expected from methods that use iso-electron density surfaces or van der Waals radii without additional empirical fitting. When using fitted radii one should use the same level of theory for the solute as was used in the parameterization (Table 1).

**Ions**. For neutral molecules solvation free energies can be measured with an accuracy of roughly 0.2 kcal/mol and reproduced theoretically to within roughly 0.5-1.0 kcal/mol, depending on the method. However, the solvation energies of ions cannot be directly measured and must be indirectly inferred relative to a standard (usually the solvation energy of the proton). The experimentally obtained solvation energies are typically accurate to within 3 kcal/mol and can be reproduced computationally with roughly the same accuracy (Kelly et al. 2006). The solvation energy of ions are therefore an especially likely source of error in binding free energies - especially if the ionic regions of the molecules become significantly desolvated due to binding.

Table 1. Common continuum solvation models used with electronic structure theory, the level of theory used for parameterization and the solvation energy of the proton used as a reference for the experimental solvation energies of ions used in the parameterization. Adapted from Ho 2015.

| Method | Level of theory used for parameterization | Solvation energy of proton used as reference for ions |
|---|---|---|
| IEFPCM-MST[a] | HF/6-31+G(d) | -264.0 kcal/mol |
| DPCM-UAHF[b] | HF/6-31(+)G(d)[c] | -261.4 kcal/mol |
| PCM-UAKS[d] | PBE1PBE/6-31G(d) | unknown |
| IEFPCM-SMD[e,f] | M05-2X98/MIDI!6D M05-2X/6-31G* M05-2X/6-31+G** M05-2X/cc-pVTZ B3LYP/6-31G* HF/6-31G* | -265.9 kcal/mol |
| COSMO-RS[g] | BP/TZVP | Not specifically parameterized for ions |
| SM8[h] | Independent of level of theory | -265.9 kcal/mol |

[a]Curutchet *et al*. 2005. IEF and CPCM give virtually identical results for water. [b]Barone *et al*. 1997. UAHF spheres have been used with CPCM with good results. [c]Diffuse functions are used only for anions. [d]This parameterization has not been published and the information is taken from the Gaussian09 manual. The method has been benchmarked for CPCM by Takano and Houk (2005). [e]Marenich *et al*. 2009. [f]The parameterization was performed by minimizing the error for all six methods simultaneously and any of the six methods can be used with the same parameter set. [g]Eckert & Klamt 2002. [h]Marenvich *et al*. 2007.

**Gas phase *vs* solution optimization**. The fitting of the radii described above is usually done using gas phase optimized structures only, i.e. any change in structure and corresponding rotational and vibrational effects are "included" in the radii via the parameterization. However, for ionic species gas phase optimization can lead to significantly distorted structures or even proton transfer and in these cases solution phase optimizations and, hence, vibrational frequency calculations, tend to be used. However, numerical noise in the continuum models can make it necessary to increase (i.e. make less stringent) the geometry convergence criteria and can lead to more imaginary frequencies than in the gas phase. One option is to compute the vibrational contribution to $\Delta G^{\circ}_{\text{gas,RRHO}}$ using gas phase optimized structures as Grimme has done (Sure *et al*. 2014).

When using solution phase geometries the gas phase single point energies needed to evaluate $\Delta G^{\circ}_{\text{solv}}(X)$ represent added computational expense one option is to use solution phase free energies to evaluate the binding free energies

$$\Delta G^\circ_{b,aq} = \Delta G^{\circ,CM}_{b,soln,E} + \Delta G^{\circ,CM}_{b,soln,RRHO} \quad (24)$$

One problem with this approach is that $\Delta G^{\circ,CM}_{b,soln,E}$, unlike $\Delta E_{gas}$, is not systematically improveable due to the empirical parameterization. For a more thorough discussion of this issue see Ho *et al*. 2010, Ribeiro *et al*. 2011 and Ho 2015.

**Cavities**. The atomic radii and corresponding cavity generation algorithms are parameterized for small molecules. For more complex molecules such as receptors this can lead to continuum solvation of regions of molecules, e.g. deep in the binding pocket, that are not accessible to the molecular solvent. Furthermore, any solvent molecule inside such pocket is likely to be quite "un-bulk-like" and not well-represented by the bulk solvent or fixed by the underlying parametrization. However, how big an error this may introduce to the binding free energy is not really known, but certain models for the cavitation energy have been shown to give unrealistically large contributions to the binding free energy (Genheden et al. 2010; Genheden & Ryde 2012).

**Explicit water molecules**. Adding explicit solvent molecules to the receptor and/or ligand can potentially lead to more accurate results. For example, including explicit water molecules around ionic sites reduces the strong dependence of the solvation energy on the corresponding atomic radii. Also, "un-bulk-like" water molecules now are treated more naturally and the risk of solvating non-solvent-accessible regions is reduced somewhat. However, adding explicit solvent molecules increases the computational cost by increasing the CPU time needed to compute energies, perform conformational searches, and compute vibrational frequencies.

There are several approaches to include the effect of explicit solvent molecules in the binding free energy. Bryantsev *et al*. (2008) suggest computing the solvation energy by

$$G^\circ_{aq,n}(X) = G^\circ_{gas}(X) + \Delta G^\circ_{solv,n}(X) \quad (25)$$

where

$$\Delta G^\circ_{solv,n}(X) = \Delta G^\circ_{gas}(X(H_2O)_n) + \Delta G^\circ_{solv}(X(H_2O)_n) - \Delta G^\circ_{solv}((H_2O)_n) \quad (26)$$

(note that $\Delta G^\circ_{solv,0}(X) = \Delta G^\circ_{solv}(X)$) and

$$\Delta G^\circ_{gas}(X(H_2O)_n) = G^\circ_{gas}(X(H_2O)_n) - G^\circ_{gas}(X) - G^\circ_{gas}((H_2O)_n) \quad (27)$$

and

$$\Delta G^{\circ\,(liq)}_{solv}((H_2O)_n) = \Delta G^\circ_{solv}((H_2O)_n) + RT \ln([H_2O]/n) \quad (28)$$

with "o (liq)" referring to a standard state of 55.34 M (the concentration of liquid water at 298K), respectively. The term $RT\ln([H2O]/n)$ is the free energy required to change the standard state of $(H2O)_n$ from 1 M to 55.34/$n$ M.

Bryantsev *et al.* have shown that using this water cluster approach leads to a smooth convergence of the solvation free energy with respect to the cluster size n. The optimum choice of *n* is this one where an additional water molecule changes the solvation energy by less than a certain amount defined by the user. One can thereby compute the optimum number of water molecules for the receptor (*n*), ligand (*m*) and receptor-ligand complex (*l*) and then compute the change in solvation free energy as

$$\Delta\Delta G^{\circ}_{b,solv,x} = \Delta G^{\circ}_{solv,l}(RL) - \Delta G^{\circ}_{solv,n}(L) - \Delta G^{\circ}_{solv,m}(R) \quad (29)$$

and computing $\Delta E_{gas}$ and $\Delta G^{\circ}_{gas,RRHO}$ as before. One can show that this corresponds to the free energy change for this reaction

$$R(H_2O)_m(aq) + L(H_2O)_n(aq) + (H_2O)_l(liq) \rightleftharpoons RL(H_2O)_l(aq) + (H_2O)_n(liq) + (H_2O)_m(liq)$$
(Rx 4)

In principle, the free energy is zero for

$$(H_2O)_l(liq) \rightleftharpoons (H_2O)_n(liq) + (H_2O)_m(liq) + \text{sgn}(d)(H_2O)_{|d|}(liq) \quad (Rx\ 5)$$

where $d = l - m - n$ and sgn(*d*) returns the sign of *d*. So the free energy change for Reaction 4 can also be computed as the free energy change for

$$R(H_2O)_m(aq) + L(H_2O)_n(aq) \rightleftharpoons RL(H_2O)_l(aq) + \text{sgn}(d)(H_2O)_{|d|}(liq) \quad (Rx\ 6)$$

However, this is only approximately true in practice due to errors in the computed gas phase and solvation free energies. Furthermore, Reaction 6 does not really lead to any significant reduction in CPU time because the water cluster free energies only have to be computed once. However, if Reaction (6) is used then one must add an additional term correcting for the indistinguishability of water molecules

$$G^{\circ}_{gas,RRHO}(X(H_2O)_n) \rightarrow G^{\circ}_{gas,RRHO}(X(H_2O)_n) - RT\ln(n!) \quad (30)$$

and similarly for the water clusters. Using Reaction (4) leads to a cancellation of this term and also maximizes error cancellation in the other energy terms. Similar considerations apply to when using individual water molecules to the balance the reaction instead of water clusters

$$R(H_2O)_m(aq) + L(H_2O)_n(aq) \rightleftharpoons RL(H_2O)_l(aq) + d(H_2O)_{|d|}(liq) \quad (Rx\ 7)$$

One of the main reasons Reaction 4 maximizes error cancellation is that the number and type of hydrogen bonds involving water molecules are very similar on each side of the equilibrium. This can also be achieved when using Reaction 6 or 7 by ensuring that $l = m + n$, in which case the error cancellation may be comparable and will depend on the nature of the ligand, host, and water arrangement. However, Eq (30) must still be used when using Reaction 6 or 7 in this way.

When using many explicit water molecules the error in the continuum solvation energies can be reduced by ensuring that the continuum solvation energy of a single water molecule matches the experimental value of -6.32 kcal/mol at 298.15K as close as possible.

**Enthalpy and entropy contributions to the binding free energy**
It is often instructive to decompose the binding free energy into enthalpy and entropy contributions. The standard enthalpy and entropy of molecule X in aqueous solution is

$$H_{aq}^{o}(X) = E_{gas}(X) + H_{gas,RRHO}^{o}(X) + \Delta H_{solv}^{o}(X) \quad (31)$$

and

$$S_{aq}^{o}(X) = S_{gas,RRHO}^{o}(X) + \Delta S_{solv}^{o}(X) \quad (32)$$

where the standard state Eq [(3)] and symmetry correction [Eq (6)] is applied to the entropy term. Thus, in order to compute these quantities one must compute the enthalpy and entropy of solvation, which can be done by the COSMO-RS (Eckert & Klamt 2002) and SM8T (Chamberlin *et al.* 2008) solvation methods. Chamberlin *et al.* (2008) have noted that most of the temperature dependence of the aqueous solvation free energy comes from the non-polar term so simply including the effect of temperature on the dielectric constant is unlikely to give accurate results. Plata and Singleton (2015) have recently shown that $\Delta S_{solv}^{o}(X)$ can make an appreciable contribution to the energy change for reaction energies.

For a molecule (X) with $N_{conf}$ conformations the standard enthalpy and entropy is

$$H_{aq}^{o}(X) = \sum_{i=1}^{N_{conf}} H_{aq}^{o}(X_i) p(X_i) \quad (33)$$

and

$$S_{aq}^{o}(X) = \sum_{i=1}^{N_{conf}} S_{aq}^{o}(X_i) p(X_i) - R \sum_{i=1}^{N_{conf}} p(X_i) \ln(p(X_i)) \quad (34)$$

where

$$p(X_i) = \frac{e^{-\Delta G^o_{aq}(X_i)/RT}}{\sum_{i=1}^{N_{conf}} e^{-\Delta G^o_{aq}(X_i)/RT}} \quad (35)$$

and $\Delta G^o_{aq}(X_i)$ is computed relative to the conformation with the lowest free energy.

The Legendre transformed entropy and enthalpy is

$$S'^o_{aq}(X_i) = -\left(\frac{\partial G'^o_{aq}(X_i)}{\partial T}\right)_{p,pH} = S^o_{aq}(X_i) - n_i(H^+)\left(\Delta S^o_{solv}(H^+) + R\ln(10)pH\right) \quad (36)$$

and

$$H'^o_{aq}(X_i) = H'^o_{aq}(X_i) - n_i(H^+)\Delta H^o_{solv}(H^+) \quad (37)$$

When comparing computed enthalpy and entropy changes to experimental measurements on systems with ionizable groups note that the observed values will depend on the buffer used *if* protonation states change upon binding (see e.g. Dullweber et al. 2001). Unless the experimental study has corrected for this effect by repeating the measurements in different buffers, this effect can contribute to the difference between the computed and experimental values.

**A Concrete Example**
In this section I apply the key equations discussed above to a specific example: *p*-xylylenediamine (L, Figure 2) binding to CB7 (R) for which a binding free energy of -9.9 ± 0.1 kcal/mol has been measured at pH 7.4 and 298K (Muddana *et al*. 2014). The conformations and other details such as the number of water molecules are just selected and constructed *for illustration purposes only* using the Avogadro program (Hanwell et al. 2012) and the MMFF force field and should not be considered accurate.

CB7 has one conformation with $D_{7h}$ symmetry and no ionizable groups. It is assumed that the solvation energy can be computed accurately without explicit water molecules. Thus, the free energy is aqueous solution is

$$G^o_{aq}(R) = G^o_{aq,0}(R) = G^o_{gas}(R) + \Delta G^o_{solv}(R) + RT\ln(14) \quad (38)$$

where $G^o_{gas}(R)$ is computed in $C_1$ symmetry and 14 is the symmetry number (σ) corresponding to the $D_{7h}$ point group.

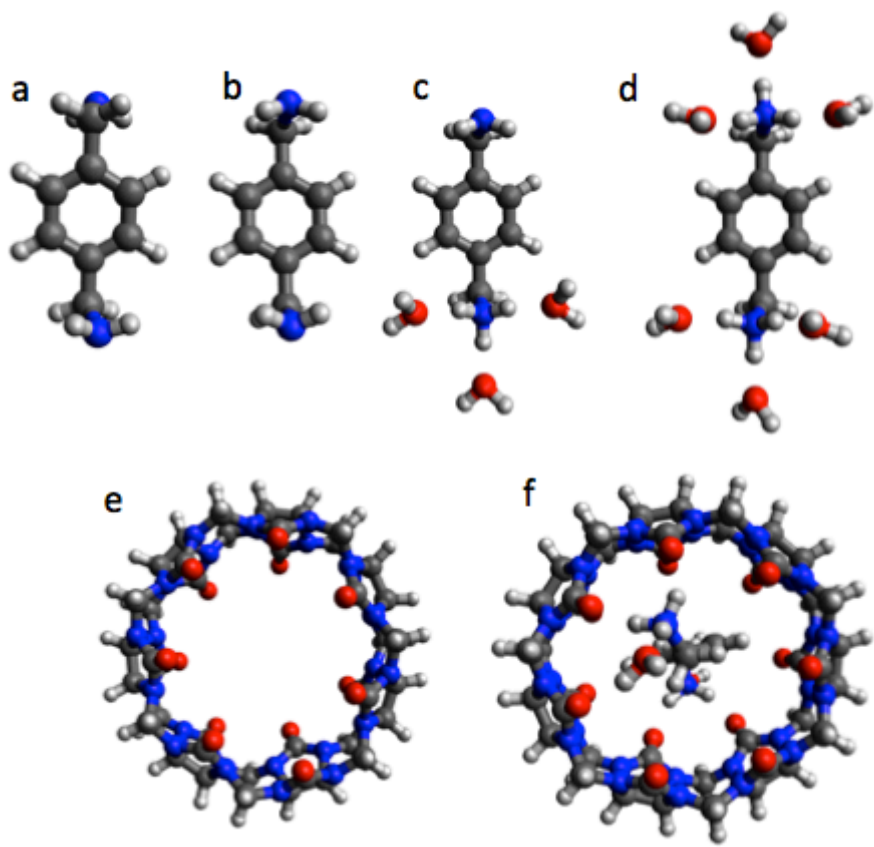

**Figure 2**. Representative conformations of ligand L (*p*-xylylenediamine), receptor R (CB7), and a receptor-ligand complex RL used to illustrate the use of the equations presented in this paper. (a) L*a*, (b) L*b*, (c) LH$^+$*b*, (d) LH$_2^{2+}$*b*, (e) R, and (f) RLH$_2^{2+}$*a*. The coordinates for the structures are available here:
http://dx.doi.org/10.6084/m9.figshare.1290639

Ligand L has two basic groups and is assumed to have two conformations *a* and *b* for each protonation state. The p$K_a$ values for the basic groups are 9.2 and 9.8 according to Marvin, so both groups are likely 100% protonated at pH 7. However, for illustration purposes I will include all three protonation states in the computation of the free energy. Furthermore, I will assume that each charged amine group is microsolvated by three explicit water molecules.

The free energy of conformer *a* of the doubly protonated state (LH$_2^{2+}$) is thus

$$G^o_{aq,6}(LH_2^{2+}a) = G^o_{gas}(LH_2^{2+}(H_2O)_6 a) + \Delta G^o_{solv}(LH_2^{2+}(H_2O)_6 a) \\ - G^o_{gas}((H_2O)_6) - \Delta G^o_{solv}((H_2O)_6) - RT\ln([H_2O]/6) - RT\ln(2) \quad (39)$$

where the gas phase energy is computed in $C_1$ symmetry and 2 is the symmetry number of the $C_2$ point group. The lowest energy structure of (H2O)$_6$ suggested by Bransyev et al. can be used for compute $G^o_{aq,n}((H_2O)_6)$, or the effect of additional conformations can be included using Eq (7). Finally, the Legendre transformed free energy [Eq (15)] at pH 7 is computed by

$$G'^o_{aq,6}(LH_2^{2+}a) = G^o_{aq,6}(LH_2^{2+}a) - 2\left(\Delta G^o_{solv}(H^+) - RT\ln(10)\text{pH}\right) \quad (40)$$

The corresponding free energy of conformer b, $G'^o_{aq,6}(LH_2^{2+}b)$, which has $C_{2v}$ symmetry and for which $\sigma$ is also 2, is computed in the same way. Notice that each conformation in principle can have different numbers of water associated with them. Similarly, the free energies of the singly protonated and neutral ligand (with $C_1$ and $C_2$ symmetry) is computed by

$$\begin{aligned} G'^o_{aq,3}(LH^+a) = & \; G^o_{gas}(LH^+(H_2O)_3 a) + \Delta G^o_{solv}(LH^+(H_2O)_3 a) \\ & - G^o_{gas}((H_2O)_3) - \Delta G^o_{solv}((H_2O)_3) - RT\ln([H_2O]/3) \\ & - \left(\delta G^o_{solv}(H^+) - RT\ln(10)\text{pH}\right) \end{aligned} \quad (41)$$

and

$$G^o_{aq,0}(LHa) = G^o_{gas}(La) + \Delta G^o_{solv}(La) + RT\ln(2) \quad (42)$$

(here for conformer a and similarly for conformer b). Finally, the free energy of L averaged over conformations and protonation states is

$$G'^o_{aq,x}(\bar{L}) = G^o_{aq,0}(La) - RT\ln\left(1 + e^{-\Delta G^o_{aq,0}(Lb)/RT} + e^{-\Delta G'^o_{aq,3}(LH^+a)/RT} + e^{-\Delta G'^o_{aq,3}(LH^+b)/RT}\right. \\ \left. + e^{-\Delta G'^o_{aq,6}(LH_2^{2+}a)/RT} + e^{-\Delta G'^o_{aq,6}(LH_2^{2+}b)/RT}\right) \quad (43)$$

where

$$\Delta G^o_{aq,0}(Lb) = G^o_{aq,0}(Lb) - G^o_{aq,0}(La) \quad (44)$$

and similarly for the remaining terms in the sum. Notice that for each conformation there are three protonation states rather than ($2^2$) because the two singly protonated structures are equivalent.

For the host-guest complex I have assumed that each conformation can bind CB7 in only one way and that two explicit water molecules per protonated group is lost upon binding, so that

$$G'^{o}_{aq,x}(R\bar{L}) = G^{o}_{aq,0}(RLa) - RT\ln\left(1 + e^{-\Delta G^{o}_{aq,0}(RLb)/RT} + e^{-\Delta G'^{o}_{aq,1}(RLH^+a)/RT} + e^{-\Delta G'^{o}_{aq,1}(RLH^+b)/RT}\right.$$
$$\left. + e^{-\Delta G'^{o}_{aq,2}(RLH_2^{2+}a)/RT} + e^{-\Delta G'^{o}_{aq,2}(RLH_2^{2+}b)/RT}\right)$$
(45)

Note that the effect of the 28 equivalent binding modes to other oxygen atoms for e.g. $LH_2^{2+}a$ (Figure 2f) is accounted for by the symmetry factors. Finally, the binding free energy is computed using Eq (16).

**Protein-Ligand Binding**
In order for the electronic structure approach to be used in drug design corresponding calculation have to be carried out on proteins, which are significantly larger than the hosts that have been used to benchmark the approach so far. QM/MM is of course the obvious choice for computing the geometries and gas phase energies, although linear scaling all QM methods such as the FMO (Fedorov *et al*. 2012) method is also possible. Furthermore, continuum methods such as PCM have been adapted for large systems and interfaced to both QM/MM (Li et al. 2003) and the FMO method (Fedorov *et al*. 2006). Of course as the system size increases conformational sampling will become a bigger practical issue.

The main issue is the computation of vibrational frequencies for the protein and protein-ligand complex. The fast semi-empirical methods currently used for computing the vibrational frequencies (dispersion and hydrogen bond-corrected PM6 and DFTB as well as HF-3c) must be made interfaced with QM/MM codes and/or be implemented in a linear scaling approach that allow for frequency calculations. Dispersion-corrected PM6 and DFTB are already implemented in AMBER, a FMO implementation of DFTB has recently been added to GAMESS (Nishimoto et al. 2014) and a similar HF-3c/FMO implementation is forthcoming from my lab.

Most QM/MM studies of enzyme catalysis constrain the geometry of a significant portion of the system to avoid spurious structural fluctuation far away from the active site contributing to the barrier. This may well be necessary for binding free energy calculations as well, in which case the effect of the constraints on the vibrational frequencies must be accounted for (Ghysels et al. 2007). Alternatively, only the Hessian of the un-constrained region can be computed (Li & Jensen 2002).

So while there is some code-adjustment to be done it may well be that the promising developments in electronic structure-based prediction of aqueous binding free energies may also be brought to bear on drug design within the next few years.

**Summary and outlook**
Recent predictions of absolute binding free energies of host-guest complexes in aqueous solution using electronic structure theory have been encouraging for some systems. It is interesting to consider the underlying innovations that have lead to the recent increase in accuracy in predicted binding free energies. Advances in computer hardware and coupled cluster algorithms made it possible to construct benchmark sets of accurate electronic binding energies for a diverse set of molecules. These benchmarks sets were then used to develop the dispersion corrections needed for accurate DFT-based electronic binding energies and the short-range (hydrogen bond) corrections to the semi-empirical methods needed to compute accurate vibrational frequencies for the RRHO free energy corrections.  In fact methods like HF-3c (Sure & Grimme 2013), while containing empirical corrections, was developed without reference to any experimental data. Another interesting observation is that the dispersion and RRHO free energy contributions to the binding free energy have roughly the same magnitude, but opposite signs. So including just one of the corrections is likely to significantly increase the error relative to experiment and lead to the wrong conclusions regarding their importance.

While there have been reasonably accurate predictions for some host-guest systems, other systems remain problematic. In paper I summarize some of the many factors that could easily contribute 1-3 kcal/mol at 298 K: three-body dispersion effects, molecular symmetry, anharmonicity, spurious imaginary frequencies, insufficient conformational sampling, wrong or changing ionization states, errors in the solvation free energy of ions, and explicit solvent (and ion) effects that are not well-represented by continuum models.

While I focus on binding free energies in aqueous solution it is worth noting that the approach also applies to any free energy difference in solution, such as conformational and reaction free energy differences or activation free energies.  Furthermore, the equations apply to solvents other than water as long as the concentration of liquid water, the solvation free energy of the proton changed, and the parameterization of the continuum solvation model are changed to match the solvent of interest. Furthermore, while the recent successes with electronic structure-based approaches have been for host-guest complexes they can be extended to protein-ligand complexes with a few methodological improvements (mainly related to the computation of vibrational frequencies).  Thus, it may well be that the promising developments in electronic structure-based prediction of aqueous binding free energies may also be brought to bear on drug design within the next few years

**Acknowledgement**
I thank the following people for very illuminating discussions and/or comments on the manuscript: Vyacheslav Bryantsev, Chris Cramer, Mike Gilson, Stefan Grimme, Fahmi Himo, Junming Ho, Adrian Jinich, Andreas Klamt, Pedro Silva, Dan Singleton and Casper Steinmann.